\documentclass{ws-procs975x65}

\begin{document}

\title{SIMULATIONS OF SOLAR SYSTEM OBSERVATIONS\\IN ALTERNATIVE THEORIES OF GRAVITY}

\author{A. HEES$^{1,2,3,*}$, B. LAMINE$^4$, S. REYNAUD$^4$, M.-T. JAEKEL$^5$, C. LE PONCIN-LAFITTE$^2$, V. LAINEY$^6$, A. F\"UZFA$^3$, J.-M. COURTY$^4$, V. DEHANT$^1$, P. WOLF$^2$}

\address{$^1$ Royal Observatory of Belgium (ROB), Avenue Circulaire 3, 1180 Bruxelles, Belgium\\
$^2$ LNE-SYRTE, Observatoire de Paris, CNRS, UPMC, 75014 Paris, France\\ 
$^3$ Namur Center for Complex System (naXys), University of Namur, Belgium\\ 
$^4$ Laboratoire Kastler Brossel, \'Ecole Normale Sup\'erieure, CNRS, UPMC, Paris, France\\
$^5$ Laboratoire de Physique Th\'eorique de l'\'Ecole Normale Sup\'erieure, CNRS, UPMC, Paris\\
$^6$ IMCCE, Observatoire de Paris, UMR 8028 du CNRS, UPMC, Universit\'e de Lille 1
}

\begin{abstract}
In this communication, we focus on the possibility to test General Relativity (GR) with radioscience experiments. We present simulations of observables performed in alternative theories of gravity using a software that simulates Range/Doppler signals directly from the space time metric. This software allows one to get the order of magnitude and the signature of the modifications induced by an alternative theory of gravity on radioscience signals. As examples, we present some simulations for the Cassini mission in Post-Einsteinian gravity (PEG) and with Standard Model Extension (SME).
\end{abstract}


\bodymatter

\section{Introduction}
There is still a great interest in testing GR (theoretical motivations such as quantification of gravity, unification with other interactions\dots). Within the solar system, the gravitational observations are always related with radioscience measurements (Range and Doppler) or with angular measurements (position of body in the sky, VLBI). In this communication, we present simulations of radioscience observables in alternative theories of gravity. The software used (presented in Hees et al~\cite{hees:2012fk}) performs simulations directly from space-time metric allowing to consider a wide class of alternative theories of gravity. The main idea is to perform a simulation in a particular alternative theory of gravity and then to adjust a GR model to the simulated data (which corresponds to fit  the initial conditions of bodies, the mass of planets, \dots). The residual signal obtained after the fit is the incompressible signature due to the alternative theory of gravity. This signature corresponds to residuals that would be obtained by a naive observer measuring data and analyzing them in GR (using standard procedure) while the correct gravitation theory is the considered alternative theory. These signatures are characteristic from the alternative theories of gravity and should be searched in residuals of real data analysis. The comparison of the amplitude of these signatures with the accuracy of the measurements gives an estimation of the uncertainties on the parameters characterizing the theory that would be reachable in a real data analysis. Furthermore, under the assumption that no anomalous residual with amplitude larger than the measurement accuracy was observed in real data analysis, we can obtain an estimation of an upper limit on parameters entering the alternative theory of gravity.

In this communication, we present radioscience simulations of Cassini spacecraft during its cruise from Jupiter to Saturn with Post-Einsteinian modifications of gravity (PEG)~\cite{jaekel:2005zr,jaekel:2006kx} and with Standard Model Extension (SME) in the gravitation sector~\cite{kostelecky:2004fk,bailey:2006uq}.

\section{Post-Einsteinian Gravity (PEG)} 
From a phenomenological point of view this theory consists in including two potentials $\Phi_N(r)$ and $\Phi_P(r)$ to the metric~\cite{jaekel:2005zr,jaekel:2006kx}. Here, we consider a series expansion of these potential, which means the metric can be written as
\begin{equation}
     g_{00}(r) = \left[g_{00}\right]_{GR}+2\delta \Phi_N(r), \qquad       g_{ij}(r) =  \left[g_{ij}\right]_{GR}+2\left(\delta \Phi_N(r) -\delta\Phi_P(r) \right)\delta_{ij}
\end{equation}                                                 
with
\begin{equation}
	\delta\Phi_N=\alpha_1 r+\alpha_2r^2, \qquad \delta\Phi_P=\chi_1r+\chi_2r^2+\delta\gamma\frac{GM}{c^2r}
\end{equation}
where $M$ is the central mass, $G$ the gravitational constant, $c$ the speed of light, $\delta\gamma=\gamma-1$ is the PPN parameter and $\alpha_i$ and $\chi_i$ are PEG parameters. 

The incompressible signatures due to the 5 PEG parameters ($\chi_i$, $\alpha_i$ and $\delta\gamma$) are represented on \fref{figSig} (a). These signatures should be searched in residuals of real data analysis. The comparison of the amplitude of these signatures with the Cassini Doppler accuracy ($10^{-14}$) gives the uncertainties on the PEG parameters reachable in a real data analysis. These uncertainties are given on \tref{tab} (a). With the assumption that no residuals larger than $10^{-14}$ has been observed in the analysis of real data, they can be interpreted as an estimation of an upper bound on the PEG parameters.

\def\figsubcap#1#2{\par\noindent\centering\footnotesize(#1): #2}          
\begin{figure}[htb]%
\begin{center}
\parbox{0.495\linewidth}{ \includegraphics[width=\linewidth]{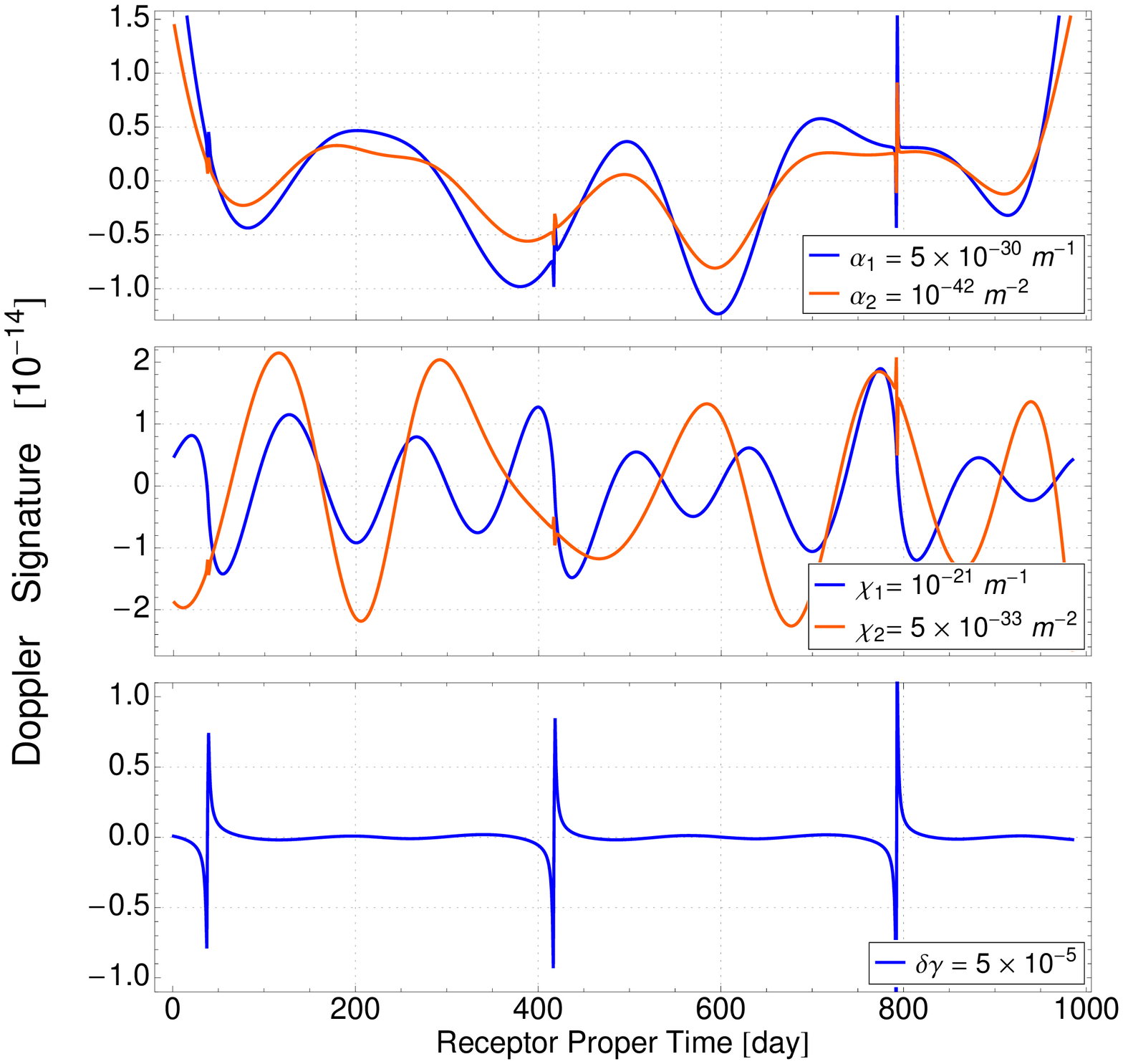}%\epsfig{figure=combined.eps,width=2in}
\figsubcap{a}{PEG}}
\hfill
\parbox{0.495\linewidth}{\includegraphics[width=\linewidth]{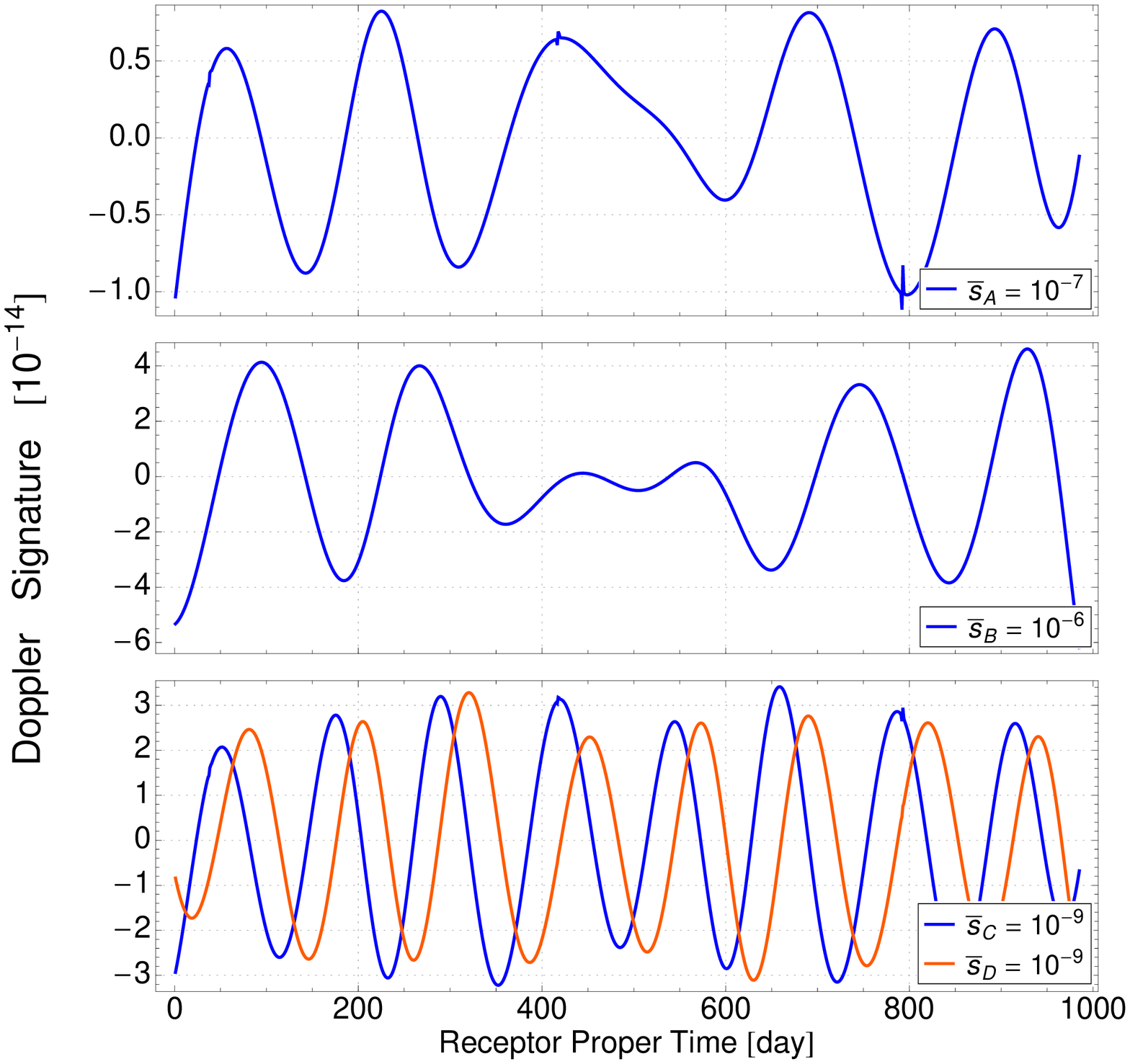} %\epsfig{figure=epsilon.eps,width=2in}
\figsubcap{b}{SME}}
\caption{Representation of the incompressible signatures due to PEG or SME theory on Doppler signal between Earth and Cassini probe.}
\label{figSig}
\end{center}
\end{figure}

\def\tabsubcap#1#2{\noindent\centering\footnotesize(#1): #2\\\vspace{1mm}} 
\begin{table}
\tbl{Uncertainties reachable in estimation of parameters.}
{
\parbox[t]{0.495\linewidth}{\begin{center}\tabsubcap{a}{PEG} \begin{tabular}{cc}
\toprule
Parameters & Reachable uncertainties\\\colrule
$\alpha_1$ & $1.9 \times 10^{-30} \ m^{-1}$\\
$\alpha_2$ & $6.2 \times 10^{-43} \ m^{-2}$ \\ 
$\chi_1$ & $5.3 \times 10^{-22} \ m^{-1}$ \\
$\chi_2$ & $1.9 \times 10^{-33} \ m^{-2}$ \\
$\delta\gamma$ &  $3.7 \times 10^{-5\phantom{0}} \ \phantom{m^{-2}}$ \\\botrule
\end{tabular}\end{center}}
\parbox[t]{0.495\linewidth}{\begin{center}\tabsubcap{b}{SME}\begin{tabular}{cc}
\toprule
Parameters & Reachable uncertainties \\\colrule
$\bar s_A$ & $8.96 \times 10^{-8\phantom{0}}$\\
$\bar s_B$ & $1.62 \times 10^{-7\phantom{0}}$ \\ 
$\bar s_C$ & $2.93 \times 10^{-10}$ \\
$\bar s_D$ & $3.05 \times 10^{-10}$ \\\botrule 
\end{tabular}\end{center}} 
\label{tab}}
\end{table}

\section{Standard Model Extension (SME)}
SME is a very general framework aimed at considering violations of Lorentz invariance. In this communication, we focus only on the gravitational sector of SME (modification in the matter sector is let for further work). In the gravitational sector of SME, the Post-Newtonian metric depends on a symmetric trace-free tensor $\bar s_{\mu\nu}$ \cite{bailey:2006uq,tso:2011uq}. We have shown that the radioscience measurements of the Cassini arc depends on 4 linear combinations of these 9 fundamental parameters:
\begin{subequations}
	\begin{align}
		\bar s_A & = \bar s_{TX}, \qquad  \bar s_B  = \bar s_{TZ}  +2.24\bar s_{TY} \\
		\bar s_C&=\bar s_{XZ}+2.31 \bar s_{XY},\\
		\bar s_D &= \bar s_{YZ} + 0.22 \bar s_{ZZ} +1.16 \bar s_{YY}-1.37\bar s_{XX}.
	\end{align}
\end{subequations}  
The incompressible signatures due to these 4 parameters are represented on \fref{figSig}~(b). These signatures should be searched in residuals of real data analysis. The uncertainties on the parameters reachable by an analysis of real data are indicated on \tref{tab}~(b). These values can also be interpreted as an upper bound on these 4 parameters under the assumption that no Doppler residual larger than $10^{-14}$ has been observed in the analysis of Cassini data.

\section*{Acknowledgments}
A. Hees is supported by an FRS-FNRS Research Fellowship.
\bibliographystyle{ws-procs975x65}
\bibliography{../../../../byMe/biblio}

\begin{thebibliography}{1}

\bibitem{hees:2012fk}
A.~{Hees}, B.~{Lamine}, S.~{Reynaud}, M.-T. {Jaekel}, C.~{Le Poncin-Lafitte},
  V.~{Lainey}, A.~{F{\"u}zfa}, J.-M. {Courty}, V.~{Dehant} and P.~{Wolf}, {\em
  {Classical and Quantum Gravity}} {\bf 29}, p. 235027 (December 2012).

\bibitem{jaekel:2005zr}
M.-T. {Jaekel} and S.~{Reynaud}, {\em Classical and Quantum Gravity} {\bf 22},
  2135 (June 2005).

\bibitem{jaekel:2006kx}
M.-T. {Jaekel} and S.~{Reynaud}, {\em Classical and Quantum Gravity} {\bf 23},
  777 (February 2006).

\bibitem{kostelecky:2004fk}
V.~A. {Kosteleck{\'y}}, {\em Phys. Rev. D} {\bf 69}, p. 105009 (May 2004).

\bibitem{bailey:2006uq}
Q.~G. Bailey and V.~A. Kosteleck\'y, {\em Phys. Rev. D} {\bf 74}, p. 045001 (Aug
  2006).

\bibitem{tso:2011uq}
R.~{Tso} and Q.~G. {Bailey}, {\em Phys. Rev. D} {\bf 84}, p. 085025 (October
  2011).

\end{thebibliography}

\end{document}